\documentclass[final,5p,times,twocolumn]{elsarticle}

\usepackage{amssymb}
\usepackage{amsmath}
\usepackage{lineno}

\journal{Nuclear Instruments and Methods in Physics Research A}

\newcommand{\meter}[1]{\ensuremath{#1\,\mathrm{m}}}
\newcommand{\centimeter}[1]{\ensuremath{#1\,\mathrm{cm}}}
\newcommand{\millimeter}[1]{\ensuremath{#1\,\mathrm{mm}}}
\newcommand{\micrometer}[1]{\ensuremath{#1\,\mathrm{\muup m}}}
\newcommand{\nanometer}[1]{\ensuremath{#1\,\mathrm{n m}}}

\newcommand{\squaremeter}[1]{\ensuremath{#1\,\mathrm{m^2}}}
\newcommand{\squarecentimeter}[1]{\ensuremath{#1\,\mathrm{cm^2}}}

\newcommand{\microsecond}[1]{\ensuremath{#1\,\mathrm{\muup s}}}
\newcommand{\nanosecond}[1]{\ensuremath{#1\,\mathrm{n s}}}

\newcommand{\kilohertz}[1]{\ensuremath{#1\,\mathrm{kHz}}}
\newcommand{\megahertz}[1]{\ensuremath{#1\,\mathrm{MHz}}}

\newcommand{\pt}{\ensuremath{p_\mathrm{T}}}

\newcommand{\gevmomentum}[1]{\ensuremath{#1\,\mathrm{GeV}/c}}

\newcommand{\ppcms}[1]{\ensuremath{\sqrt{s}=#1\,\mathrm{TeV}}}
\newcommand{\pbpbcms}[1]{\ensuremath{\sqrt{s_{\mathrm{NN}}}=#1\,\mathrm{TeV}}}

\newcommand{\pblumi}[1]{\ensuremath{#1\,\mathrm{pb}^{-1}}}
\newcommand{\nblumi}[1]{\ensuremath{#1\,\mathrm{nb}^{-1}}}

\newcommand{\xpercentlayer}[1]{\ensuremath{#1\%\,X_{0}/\mathrm{layer}}}

\newcommand{\dedx}{\ensuremath{\mathrm{d}E/\mathrm{d}x} }

\newcommand{\ITS}{ITS2 }

\begin{document}


\begin{frontmatter}

\title{Performance of the ALICE Inner Tracking System 2}

\author{Nicol\`o Valle\corref{mycorrespondingauthor}}
\ead{nicolo.valle@cern.ch}
\author{on behalf of the ALICE Collaboration}
\address{INFN, Sezione di Pavia, Via A. Bassi 6, 27100 Pavia Italy}


\begin{abstract}
The upgraded Inner Tracking System (ITS2) of the ALICE experiment at the CERN Large Hadron Collider is based on Monolithic Active Pixel Sensors (MAPS). With a sensitive area of about \squaremeter{10} and 12.5 billion pixels, ITS2 represents the largest pixel detector in high-energy physics. The detector consists of seven concentric layers equipped with ALPIDE pixel sensors manufactured in the TowerJazz \nanometer{180} CMOS Imaging Sensor process. 
The high spatial resolution and low material budget, in combination with small radial distance of the innermost layer 
from the interaction point, make the detector well suited for secondary vertex reconstruction as well as for tracking at low transverse momentum.

This paper will present the detector performance during the LHC Run 3 and give an overview on the calibration methods and running experience.

\end{abstract}


\end{frontmatter}

\section{The ITS2 detector concept}\label{sec::detector}

The ALICE experiment at the LHC underwent a major upgrade during the 2019-2021 Long Shutdown, targeting high-precision studies of the quark-gluon plasma with low transverse momentum (\pt) probes. The upgrade includes a beam pipe with reduced outer radius, moved from 28 mm to 18 mm, a new Inner Tracking System (ITS2), a new Muon Forward Tracker (MFT), a new GEM-based TPC readout, and a new Fast Interaction Trigger (FIT) detector. Furthermore, the readout and trigger system has been extensively upgraded to record Pb–Pb collisions at up to 50 kHz in continuous mode, ensuring sensitivity to signals that do not have a triggerable signature.

\begin{figure}[h!]
\centering
\includegraphics[width=0.9\columnwidth]{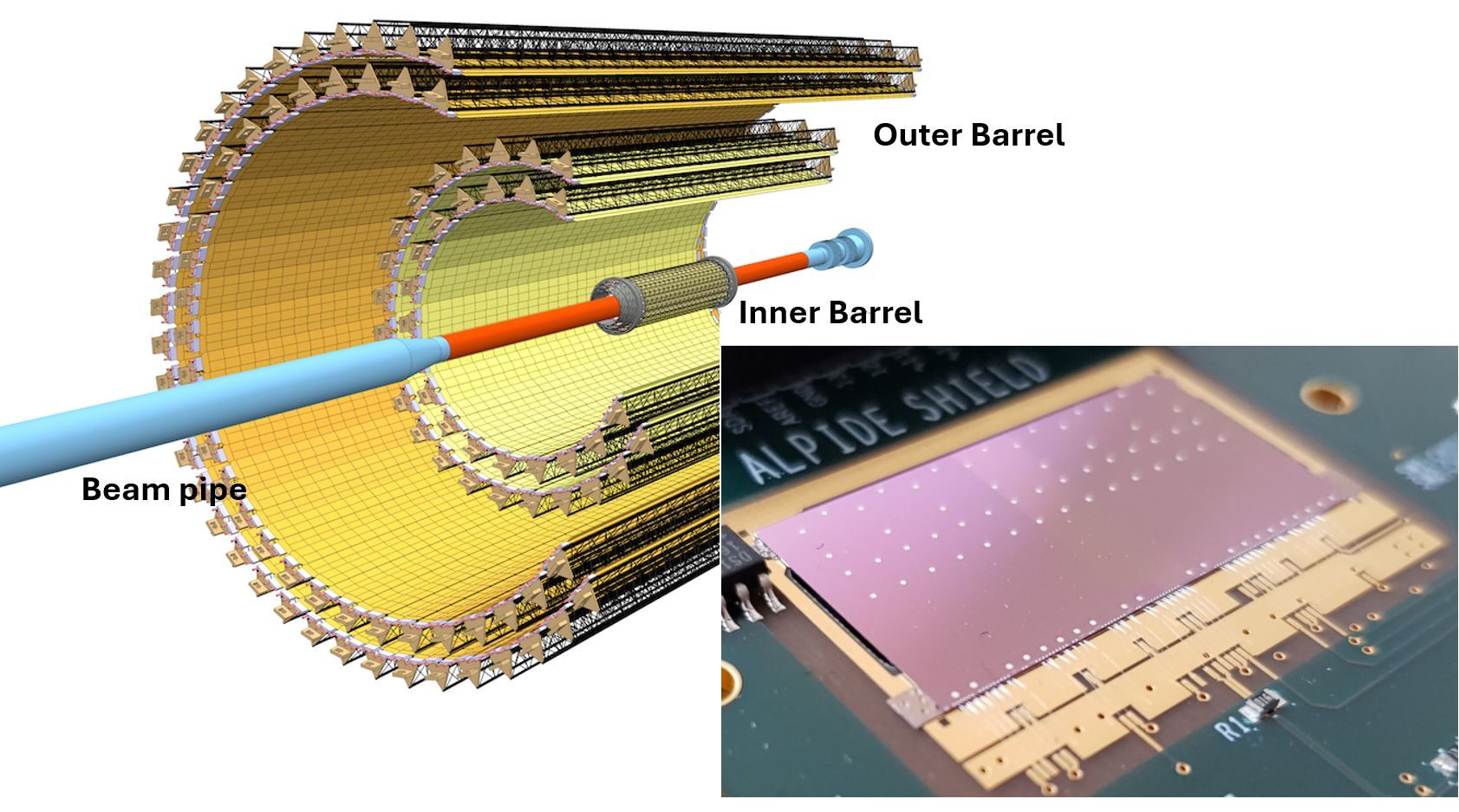}
\caption{ITS2 layout and photo of an ALPIDE chip.}\label{fig::layout}
\end{figure}

The new ITS2 
significantly enhances ALICE’s tracking capabilities, especially
at low \pt $\,$ \cite{ITS2TDR, LS2upgrade}. It consists of seven cylindrical silicon pixel layers, made of ALPIDE chips arranged in longitudinal staves as shown in Fig.~\ref{fig::layout}, for a total of 12.5 billion pixels covering a sensitive area of about \squaremeter{10}. The innermost three layers form the Inner Barrel (IB) and are installed at radii between 22 and \millimeter{40}. The outermost layer of the Outer Barrel (OB) is located at a radius of \centimeter{40}. 

The ALPIDE (ALICE Pixel Detector) chip is a $3\times \squarecentimeter{1.5}$ matrix of $1024 \times 512$ monolithic pixel sensors with $O(\micrometer{30})$ pixel pitch, reaching an intrinsic spatial resolution of about \micrometer{5}. It is  manufactured in the TowerJazz \nanometer{180} CMOS imaging process 
with a \micrometer{25} thick active layer allowing to reduce the silicon
thickness to just \micrometer{50} in the IB. Together with an ultra-light mechanical support, this allowed to achieve a material budget of \xpercentlayer{0.36} for the IB and \xpercentlayer{1.1} for the OB. Full CMOS circuitry is implemented directly into the pixel cell, which contains the sensing diode, amplification and shaping stage, discriminator,  and a digital section  with multi-event buffers and mask registers.
A pulse injection capacitor allows to inject test charges into the input of the analog front-end, a crucial feature for calibration as explained in section \ref{sec::calibration}. Additionally, the memory cells of the pixels can be directly set, to emulate hits with any desired pattern. The \ITS Detector Control System has been programmed to simulate physics events using this feature, enabling the commissioning of the entire data acquisition chain under realistic particle load conditions even without beam collisions. 

The periphery of each chip houses control and readout functionalities, along with 14 DACs for biasing the pixel front-end circuitry. These include registers for tuning the discrimination charge threshold, which can be adjusted individually for each chip. The chip matrix is read out with zero suppression using high-speed links supporting 1.2 Gb/s. 
In the OB, a slower parallel data port is used for inter-chip data transfer, allowing for the integration of compact multi-chip modules where a single high-speed link handles the readout of seven chips. Each of the 192 staves is connected to a FPGA-based Readout Unit (RU), working in the $0.5\, \mathrm{T}$ magnetic field of ALICE and in radiation environment ($10\, \mathrm{krad}$ of TID \cite{ITSReidt}).

The \ITS has been operated so far in continuous mode. The RU firmware generates periodic triggers with a programmable rate, steered by a signal received with the frequency of the LHC orbit of approximately \kilohertz{11}. Normally, a framing rate of \kilohertz{202} is used for pp data acquisition, with the ALICE interaction rate usually leveled at \kilohertz{500}. During Pb-Pb data acquisition, where the hadronic interaction rate in ALICE was pushed to \kilohertz{50} for the first time, a framing rate of \kilohertz{67} 
is used.

\vspace{0.5em}

More details on the detector system architecture and the ALICE software system can be found in \cite{ITS2TDR, LS2upgrade}. The remainder of this work focuses on the \ITS calibration methods and presents a selection of operational highlights and physics results.

\section{Calibration: charge threshold, noisy pixels and acceptance maps}\label{sec::calibration}

The optimization of the in-pixel threshold is achieved by injecting test charges into the input of the analog front-end. The process is fully managed by the \ITS Detector Control System, which triggers the chips at a custom frequency of 1 kHz using the internal sequencer equipped in the RUs, independent of the ALICE trigger. 
During a threshold scan, fifty charges (from 0 to 500 electrons) are injected fifty times each. 
The response of the detector is processed synchronously in the ALICE computing farm, where the threshold and noise levels are determined from the s-curve measured for each pixel. If tuning of the threshold is required, this procedure is repeated for a representative subset of pixels at each step during a scan of the values of the relevant on-chip biasing DACs.
Threshold values ranging from approximately 60 to 170 electrons guarantee a detection efficiency of over 99\% and a sufficiently low rate of fake hits, as shown in~\cite{LS2upgrade}. The chosen working point
is 100 $\mathrm{e}^-$.

\begin{figure}[h!]
\centering
\includegraphics[width=\columnwidth]{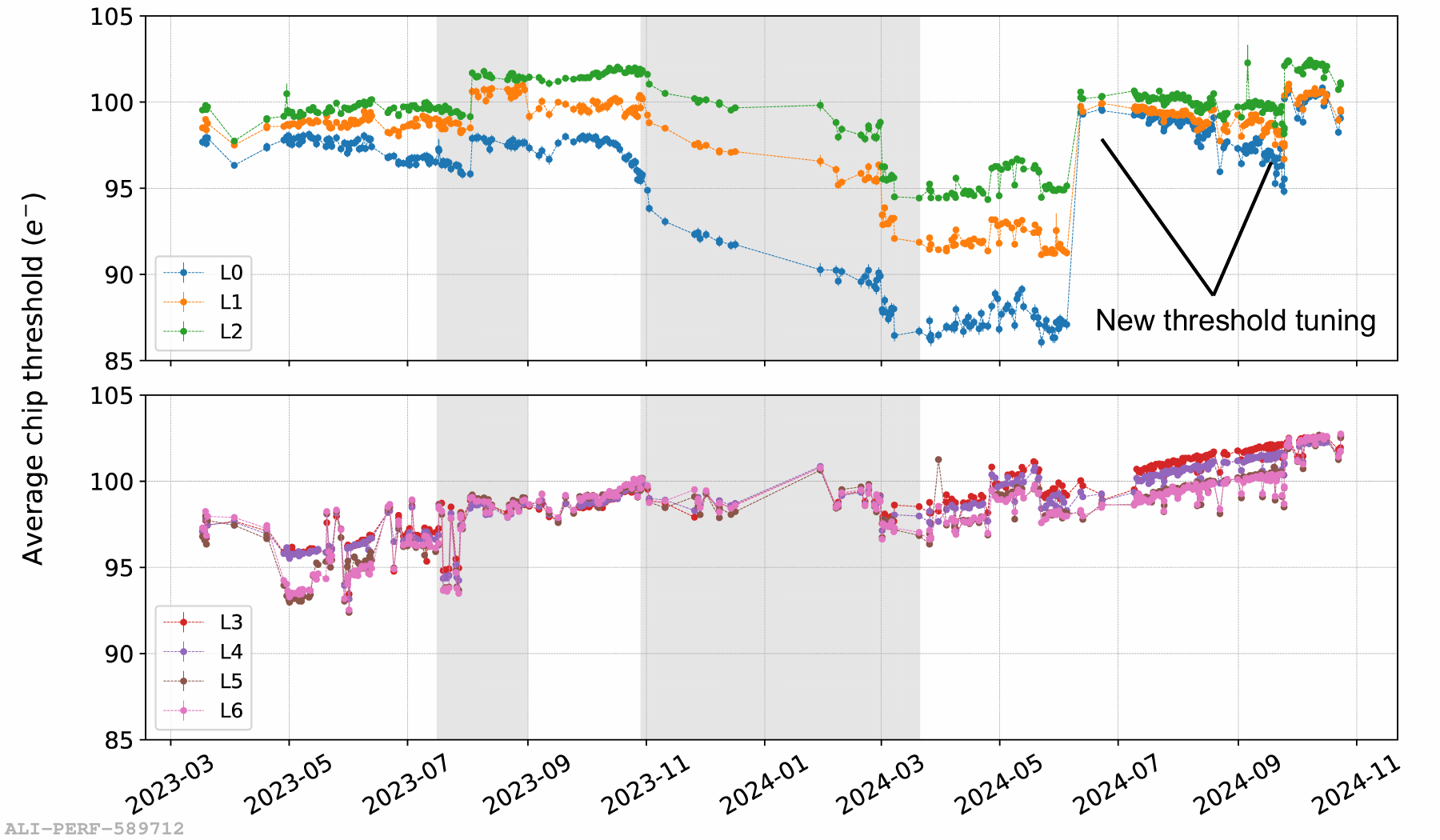}
\caption{Evolution of the average in-pixel discriminating threshold per layer, from March 2023 to October 2024. Periods without beams in the LHC are marked in gray.}\label{fig::thresholdtrend}
\end{figure}

The stability of the threshold is monitored daily. After tuning, the average chip value is uniform across the entire detector at the percent level. The stability over time is illustrated in Fig.~\ref{fig::thresholdtrend}. Small fluctuations are primarily induced by variations in the chip supply voltages. A modest shift effect is also observed in the IB and attributed to the accumulated radiation dose. Those effects are consistently recoverable with a new tuning.

After tuning the thresholds, a noise calibration is performed to identify the noisy pixels and mask them. Pixels firing more than $10^{-6}$ times per event in the OB or $10^{-2}$ times in the IB during readout without beams are classified as noisy. 
The higher threshold in the inner layers is set to prioritize the detection efficiency and remains affordable in terms of both bandwidth and event reconstruction. 
As a result of this strategy, the fraction of masked pixels is lower than $10^{-5}$ in the IB and $10^{-4}$ in the OB. 
The rate of fake hits during data taking is kept at the level of $10^{-7}$ hits/pix/evt in the IB and $10^{-8}$ hits/pix/evt in the OB, well within the detector's requirements of $10^{-6}$ hits/pix/evt \cite{ITS2TDR, O2TDR}.

\begin{figure}[h!]
\centering
\includegraphics[width=0.9\columnwidth]{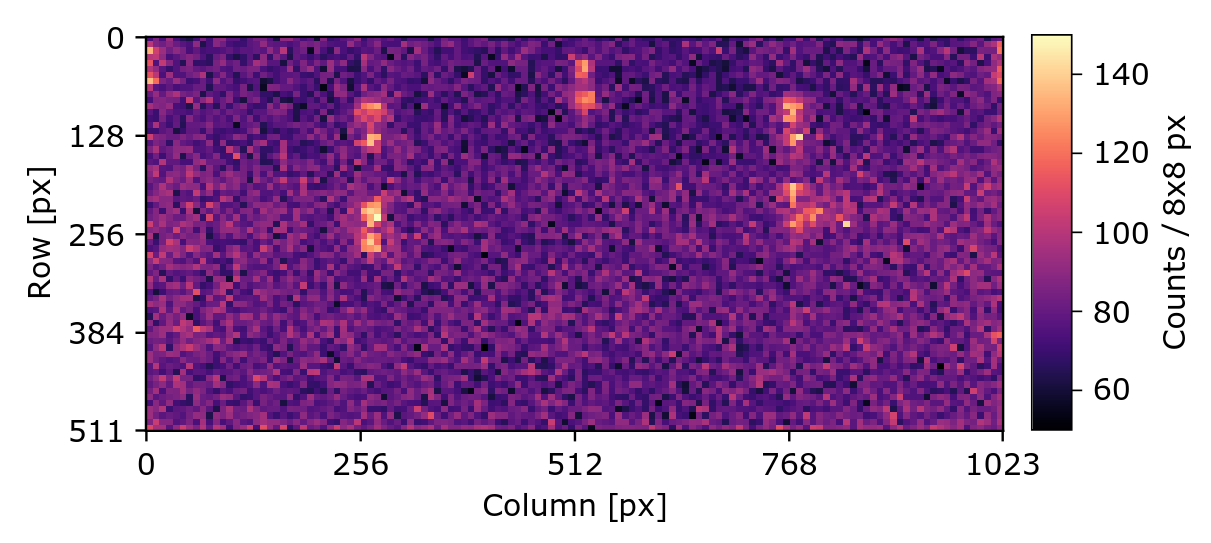}
\includegraphics[width=0.67\columnwidth]{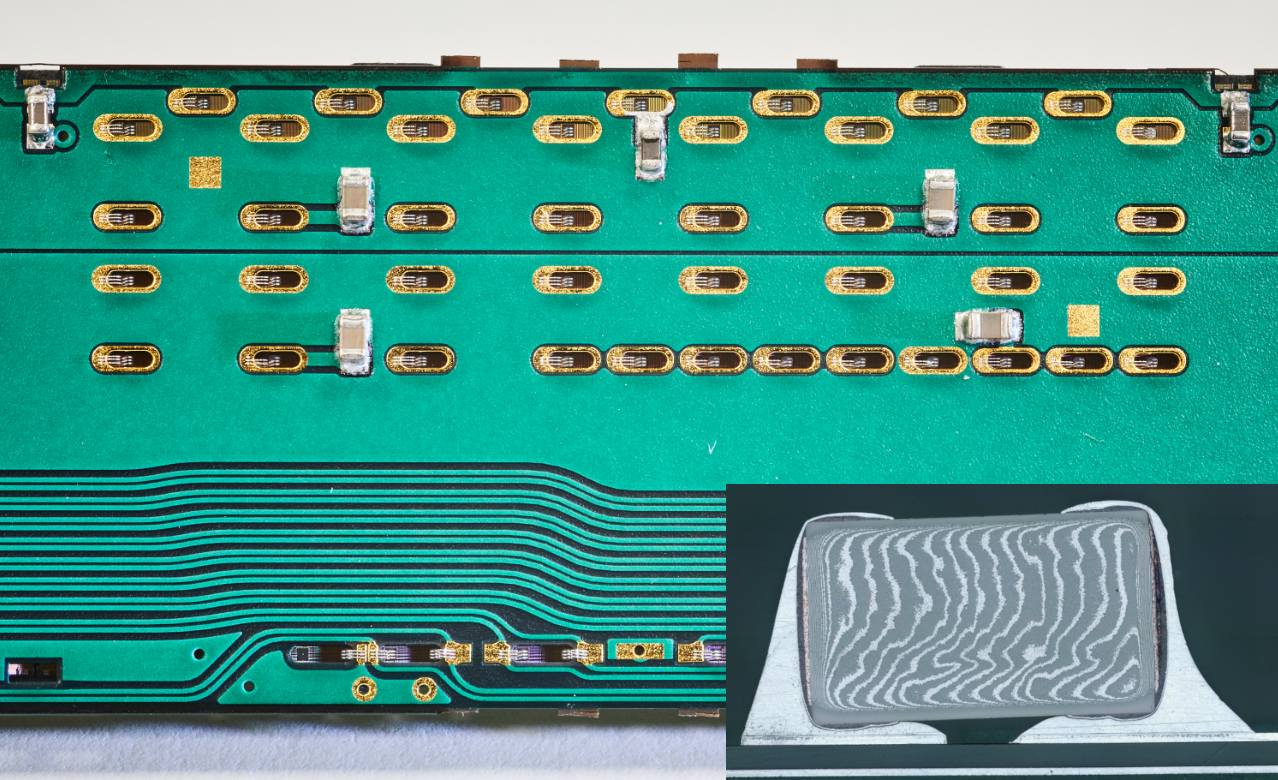}
\caption{From the on-surface commissioning. \textit{Top}: observed noise pattern across the chip. Noise maps from different chips are overlapped, giving an equivalent exposure time of 204 days. \textit{Bottom}: printed circuit glued on top of the sensor, with the cross section of one decoupling capacitor. Taken from \cite{MagerNoise}. }\label{fig::betanoise}
\end{figure}

The possibility to run with a static and loose mask was already observed during the on-surface commissioning. The top part of Fig.~\ref{fig::betanoise} shows the distribution across the chip of the residual noise observed after the application of the mask. 
Contributions from cosmic tracks and potential electronic noise were filtered out during the analysis. 
The resulting pattern matches the position of the decoupling capacitors located on the back of the printed circuit (FPC) glued to the chip (see the bottom panel of the figure). Indeed, the observed pattern was proven to be caused by the lead present in the tin soldering of the capacitor terminals. 
The $\betaup$ radiation emitted by the $\mathrm{^{210}Pb}$ isotope decaying into $\mathrm{^{210}Po}$ can penetrate the FPC and reach the chip.
This was confirmed through Geant-4 simulations and control measurements with tin samples placed on top of the chip.
More details can be found in \cite{MagerNoise}.
Such an additional hit source has no impact on the operation of the detector, however, the achieved sensibility demonstrates well the low-noise performance.

Beyond those mentioned above, a comprehensive set of calibration scans can be performed to fine-tune the detector's working point, identify dead or stuck pixels, and monitor the detector's condition. 
An additional layer of calibration is performed synchronously with data taking to detect and record time-evolving chip inefficiencies. Variations in the supply voltage, especially at the beginning of high-occupancy event readouts, can induce significant phase shifts on the high-speed links from the ALPIDE chips to the RUs, potentially causing the RU phase tracking to lose synchronization with the ALPIDE signal. This results in a loss of data from one (IB) or seven (OB) chips. The \ITS Detector Control System automatically recovers links from such occurrences. 
The resynchronization process involves suspending triggers to all the chips connected to the affected RU, for approximately 10 seconds. Data lost during these transients impact cluster and track distributions. The online calibration task uses raw data to identify missing chips and saves their addresses in acceptance maps with time granularity of at least 1 Hz. 
These maps are then used to anchor the Monte Carlo simulation to the realistic conditions of the run, ensuring reliable physics measurements.

\begin{figure}[h!]
\centering
\includegraphics[width=\columnwidth]{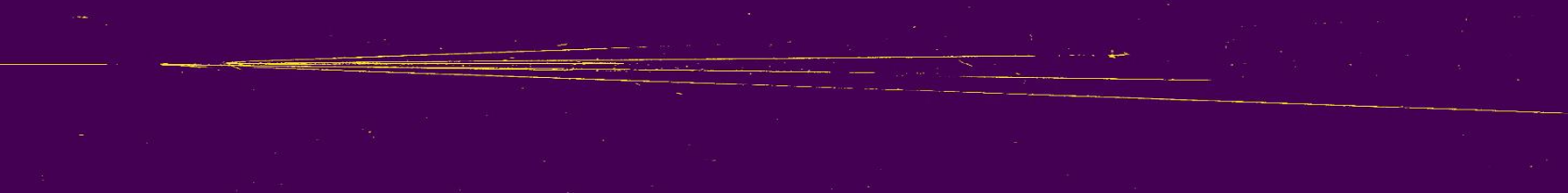}
\caption{Hitmap produced by one of the background events described in the text. The figure ($\centimeter{15} \times \centimeter{1.5})$ displays part of one stave of the innermost layer.}\label{fig::bkghitmap}
\end{figure}

It is important to note that this represents  a minor loss of acceptance. 
The average uptime is maintained at 99\% or higher. An exception to the typical performance was observed during the first days of Pb-Pb beam operation in October 2023. 
The \ITS data rate profiles, cluster occupancy, hit maps (see Fig.~\ref{fig::bkghitmap}), and subsequent analyses confirmed that particle showers (primarily muons) originating from a distant source were crossing the \ITS chips at a few centimeters from the beam pipe at a very shallow angle. 
The occupancy induced by these events and their frequency were high enough to induce a dead time of approximately 100\% on one fifth of the IB. 
The primary source of this background was identified as a $\mathrm{^{207}Pb}$ component in the beam, circulating off-trajectory and hitting a collimator located \meter{120} before the ALICE interaction region. 
The issue has been mitigated, with only \nblumi{0.2} of the \nblumi{1.96} delivered luminosity in 2023 affected by this problem.

\section{Highlights from physics data taking}\label{sec::physics}

The total luminosity delivered to  ALICE from 2022 to 2024 amounts to  $\sim\! \pblumi{82}$ in pp collisions and $\sim\! \nblumi{3.3}$ in Pb–Pb collisions. This represents roughly a factor 3000 (pp) and 60 (Pb-Pb)  more than the minimum bias samples collected in Run~1 and Run~2. The availability of the \ITS during data-taking operations has been nearly 100\%.

This section presents a selection of findings from the physics data collected by the \ITS, though it is not an exhaustive list.

\begin{figure}[h!]
\centering
\includegraphics[width=0.75\columnwidth]{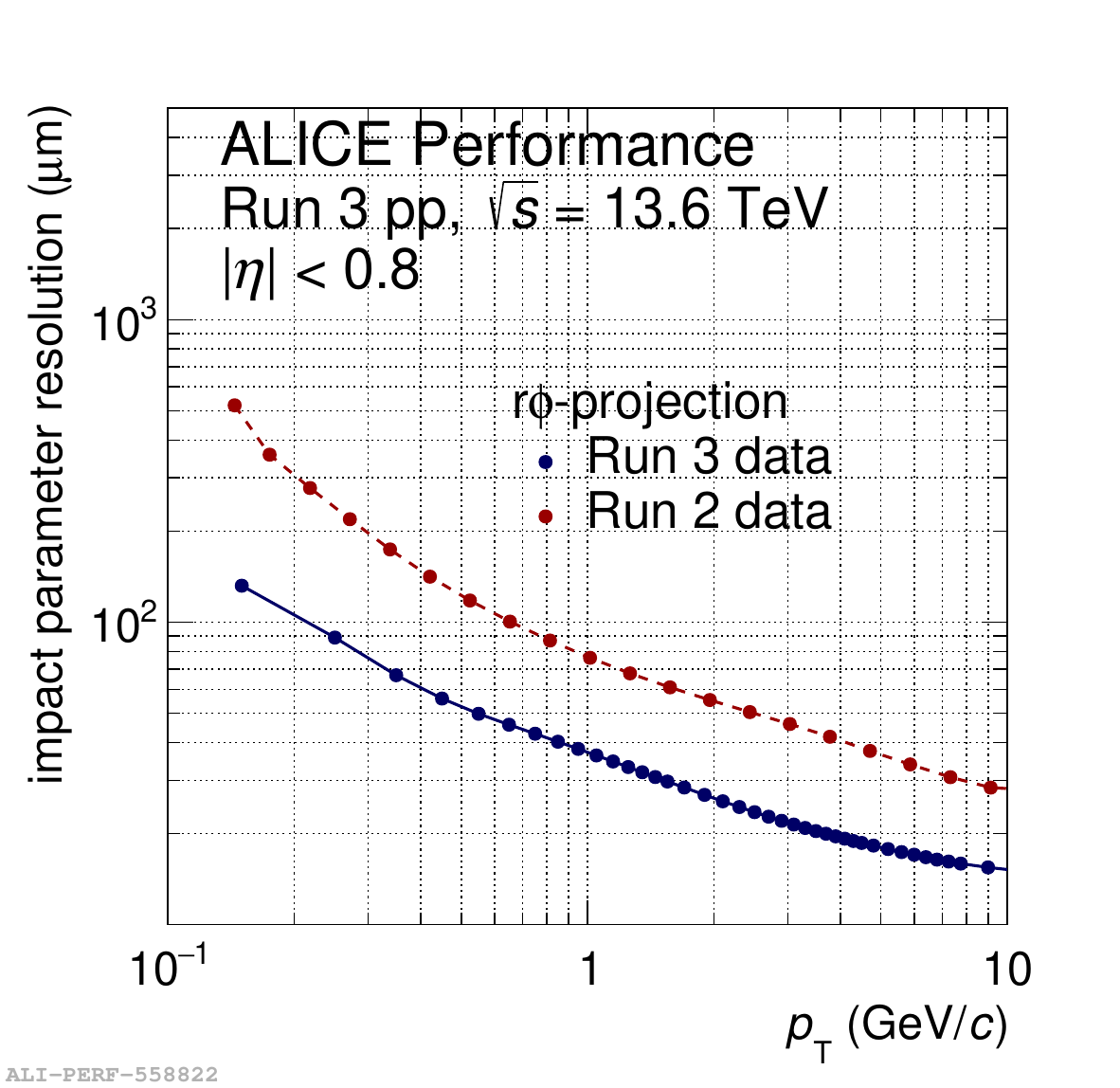}
\caption{Impact parameter resolution on the transverse plane versus \pt, measured in Run 2 pp collisions at \ppcms{13} with the ITS1 (global tracks with at least one hit in the innermost silicon pixel layers) and in Run 3 at \ppcms{13.6} with the current ITS2 (global tracks with at least one hit in the inner barrel). }\label{fig::ipres}
\end{figure}

The resolution of the track impact parameter in the transverse plane, as measured in pp collisions, is shown in Fig.~\ref{fig::ipres}. 
A comparison with Run 2 results shows a twofold improvement in the performance of the current ITS compared to the previous ITS1 detector. 

\begin{figure}[h!]
\centering
\includegraphics[width=\columnwidth]{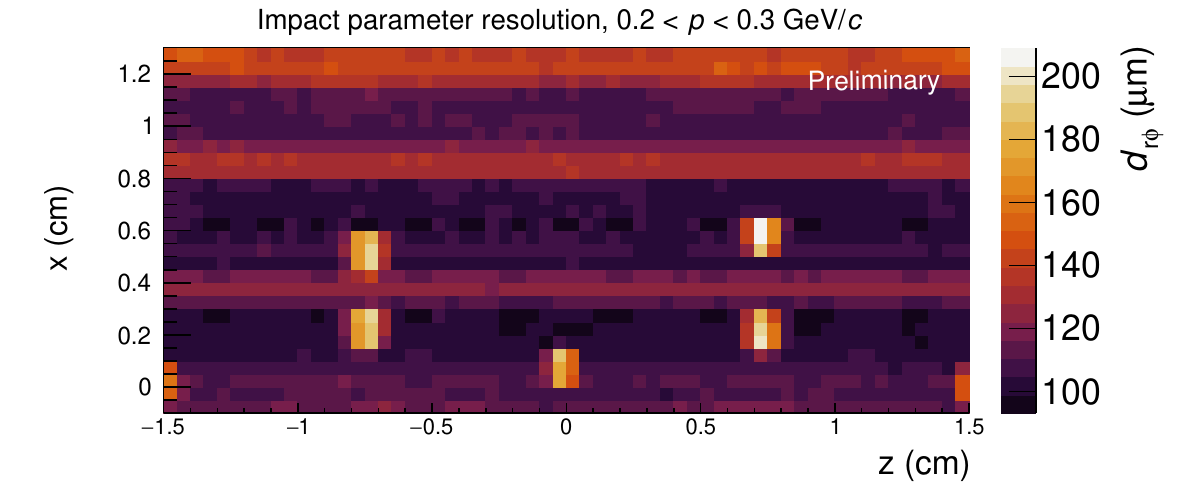}
\caption{Impact parameter resolution for tracks with $0.2 < p < \gevmomentum{0.3}$ as a function of the hit position on the central chip of one innermost stave. The regions with worse resolution match the position of the capacitors on the back of the printed circuit glued to the chip, the position of the cooling lines and the overlap of the staves at the edge of the chip.}\label{fig::samuele}
\end{figure}

Ongoing studies aim to further refine the detector alignment and material description, with the potential to enhance the pointing resolution even further. As an example, the distribution of the tracks impact parameter has been correlated to the reconstructed position on the chips. Figure \ref{fig::samuele} displays the width of the distribution as a function of the track position on the central chip ($-1.5 < z < \centimeter{1.5}$) on one of the innermost staves, for tracks with momentum $0.2 < p < \gevmomentum{0.3} $. A few regions with degraded resolution are clearly identifiable. These correspond to an increased material budget at the locations of capacitors soldered onto the FPC (see Fig.~\ref{fig::betanoise}).

\subsection{Strangeness tracking}

\begin{figure}[h!]
\centering
\includegraphics[width=0.495\columnwidth]{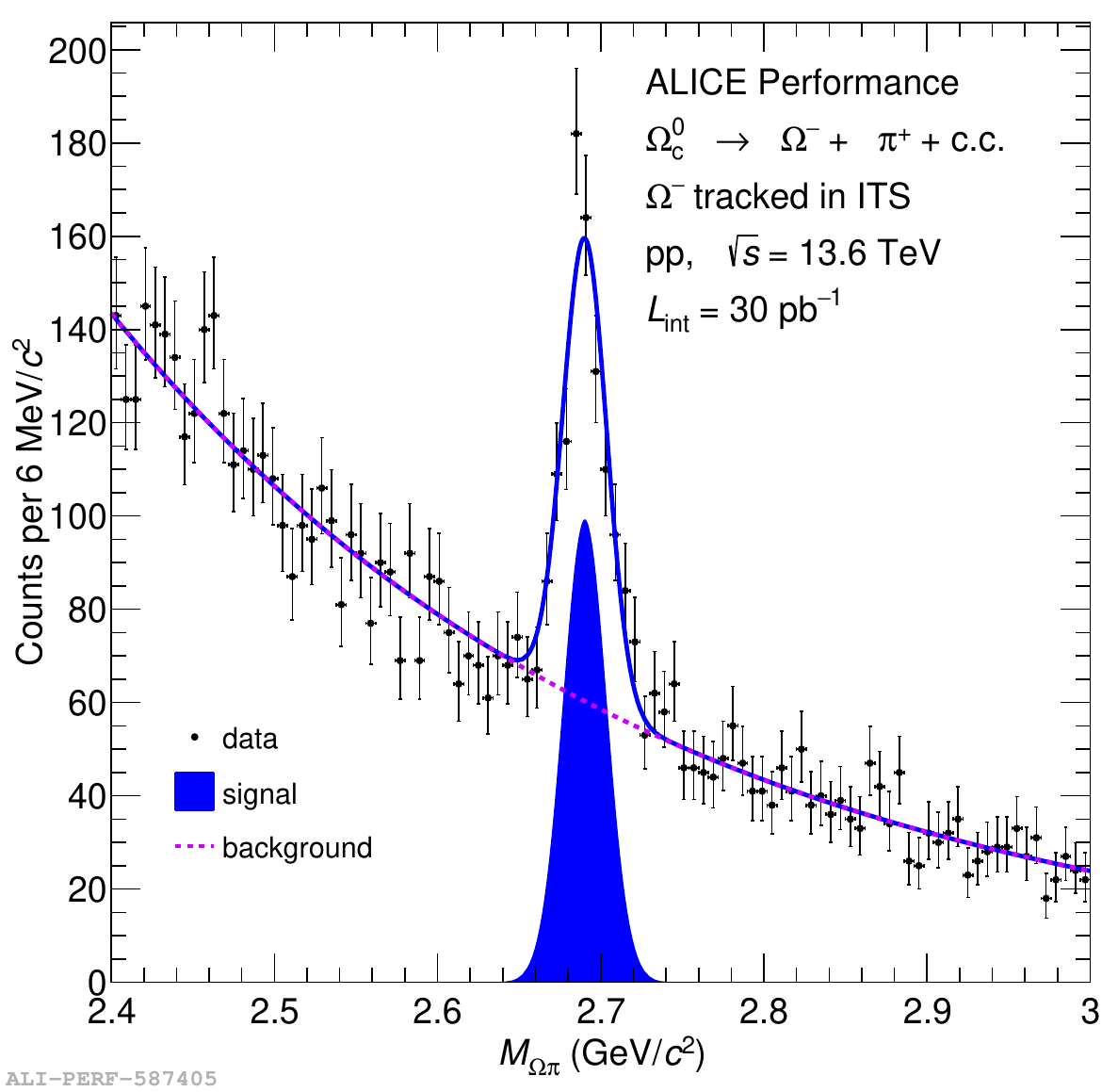}
\includegraphics[width=0.495\columnwidth]{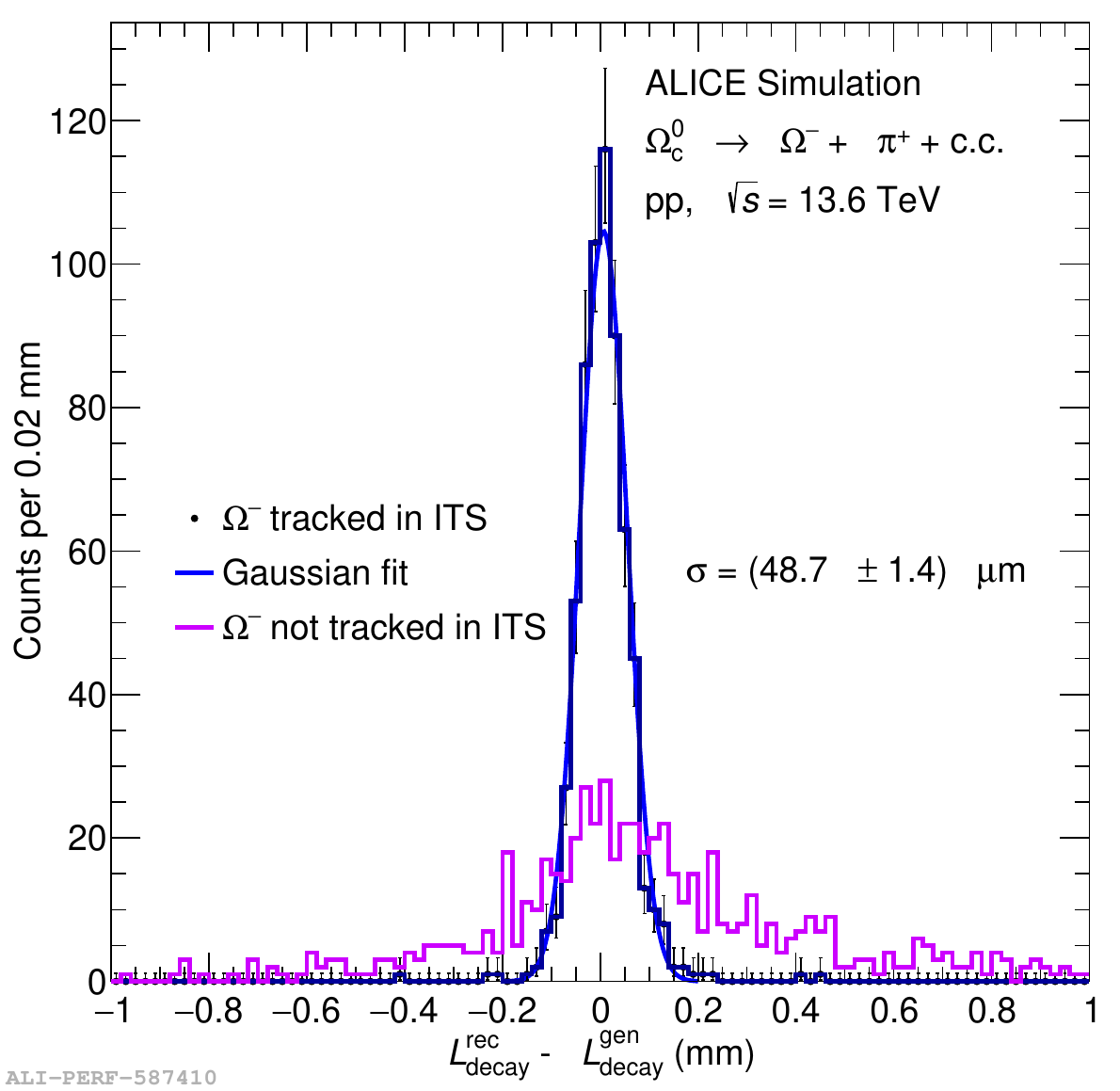}
\caption{\textit{Left}: invariant mass distribution of tracked $\Omega^\pm$ baryons paired with primary pions. The fit shows the peak from the decay $\Omega_\mathrm{c}^0\to \Omega^\pm + \piup^\mp$ and an exponential background. \textit{Right}: Simulated resolution of the decay length of $\Omega_\mathrm{c}^0$ baryons reconstructed from a pion and an $\Omega$ tracked in the \ITS.}\label{fig::omegac}
\end{figure}

Strangeness reconstruction is a key element in general-purpose detectors. Thanks to the high-resolution spatial information of the \ITS inner layers, weakly decaying particles can be detected directly. The IB tracklets are then combined with the information of topological reconstruction
from the decay daughters. This capability opens up new physical possibilities and significantly enhances background rejection and secondary vertex resolution. The measurement of mass and lifetime of charmed $\Omega_\mathrm{c}^0$ baryons is an example on how complex decay reconstruction is realized in ALICE. 
The left panel of Fig.~\ref{fig::omegac} displays the invariant mass distribution of $\Omega^- \piup^+$ (+ c.c.) pairs from pp collisions of Run 3. $\Omega^\pm$ tracked in the \ITS 
were 
paired with pion candidates to reconstruct secondary vertices, after suppressing the combinatorial background with a set of topological cuts. A fit with the peak from the $\Omega_\mathrm{c}^0\to \Omega^-+\piup^+$ decay is shown in the figure. 
Moreover, the performance of the vertexing with strangeness tracking has been proved to be good enough to measure  lifetimes. The right panel of figure \ref{fig::omegac} shows the achievable resolution on the decay length of the $\Omega_\mathrm{c}^0$ when the $\Omega^-$ daughters are tracked in the \ITS. 

\subsection{Particle identification with MAPS}

During normal data-taking operations, the analogue response of the ALPIDE is clipped to a maximum length of \microsecond{5}. This decouples the time-over-threshold (ToT) from the charge deposit, minimizing the likelihood that the signal remains above the threshold in consecutive strobe windows, while maintaining the particle detection efficiency. Deactivating the clipping allows the ToT to become proportional to the charge released in the silicon, enabling the sampling of the full signal shape by increasing the strobing rate.

This is the basis for a series of special runs conducted with the \ITS Inner Barrel. The framing rate has been increased to \megahertz{2.2}. By counting the number of consecutively readout frames with digital response, the ToT of the signal in each pixel can be measured. 
Oversampling substantially increases the volume of data, with the risk of filling the chip event buffers, causing the chip to enter a busy state that prevents it from accepting new triggers. To minimize this, the data collection was performed during dedicated pp runs at reduced interaction rate of about \kilohertz{1}.
The calibration of the measurement is done using a pulse-injection technique. For each charge injection, the delay of the ALPIDE strobe is systematically increased in steps of \nanosecond{25}. The point at which the injected charge is no longer collected corresponds to the duration that the released charge remained above the detection threshold. 

\begin{figure}[h!]
\centering

\includegraphics[width=\columnwidth]{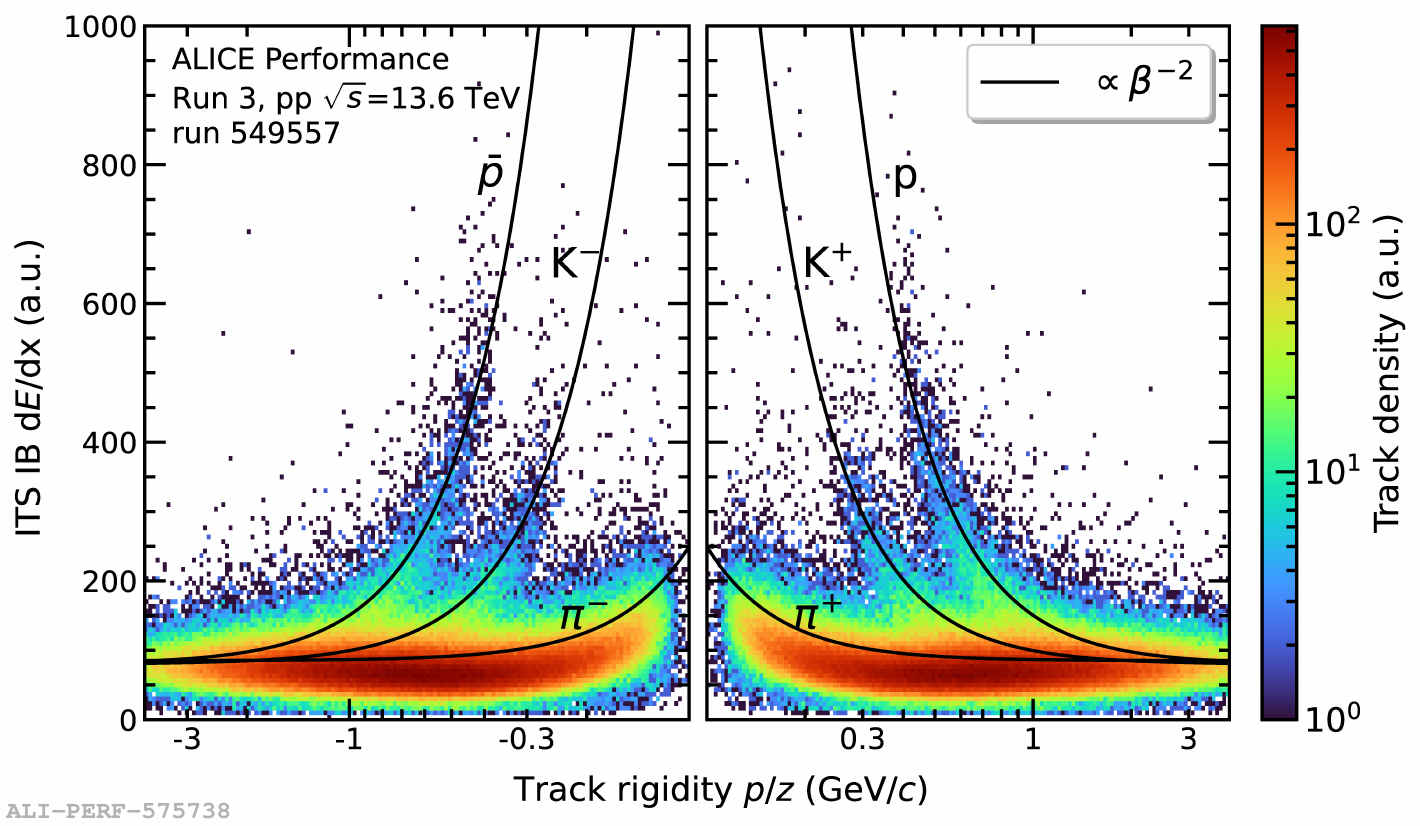}
\caption{
$\dedx$ distribution versus track rigidity measured with ITS2 Inner Barrel in pp collisions at $\pbpbcms{13.6}$.}\label{fig::dedx}
\end{figure}

Figure~\ref{fig::dedx}
presents the \dedx distribution measured in a dedicated pp run. The ToT for each pixel has been converted into charge, as described, and adjusted for the angle of incidence through the silicon to compute the energy loss. The charges from pixels within the same cluster have been summed up. \ITS tracks matching tracks in the TPC were used for analysis. For comparison, a parameterization of the energy loss as a function of $\beta^{-2} = (p/E)^{-2}$ is overlaid on the spectra for pions, kaons, and protons. This result represents the first \dedx
measurement in a MAPS pixel tracker.

 \section{Conclusions}

 The ITS2 is the ALICE's fully MAPS-based tracking system, successfully operating since the start of LHC Run 3. Key achievements during commissioning and data-taking include low noise levels, uniform and stable charge thresholds, and excellent tracking efficiency and pointing resolution.

The experience gained with the current detector has been instrumental in guiding the development of the upgraded one, ITS3, planned for installation during the next LHC Long Shutdown in 2026 \cite{KlugeVCI2025}.

\end{document}